\pdfoutput=1
\documentclass{JINST}

\title{Characterization of a scintillating fibers read by MPPC detectors trigger prototype for the AMADEUS experiment}

\author{M. Bazzi$^a$, C. Berucci$^a$, C. Curceanu$^a$, A. D'Uffizi$^a$, K. Piscicchia$^a$, M. Poli Lener$^a$, \newline A. Romero Vidal$^b$, E. Sbardella$^a$, A.Scordo$^a$\thanks{Corresponding
author.}, O. Vazquez Doce$^c$\\
\llap{$^a$}Laboratori Nazionali di Frascati, INFN,\\
  Via E. Fermi 40, Frascati (Rome), Italy\\
\llap{$^b$}Universidade de Santiago de Compostela,\
  R\'ua Xos\'e Mar\'ia Su\'arez N\'u\~{n}ez 15782 Santiago de Compostela, Spain\\
\llap{$^c$}Excellence Cluster Universe, Technische Universit\"at M\"unchen,\\
  James-Franck-Str. 1, 85747 Garching

  E-mail: \email{alessandro.scordo@lnf.infn.it}}

\abstract{Multi-Pixel Photon Counters (MPPC) consist of hundreds of micro silicon Avalanche PhotoDiodes (APD) working in Geiger mode. The high gain and the low noise, typical of these devices, together with their good performance in magnetic field, make them ideal readout detectors for scintillating fibers as trigger detectors in particle and
nuclear physics experiments like AMADEUS, where such detectors are planned to be used to trigger on charged kaon pairs. In order to investigate the detection efficiency of such a system, a prototype setup consisting of 32, 1 mm diameter scintillating fibers, arranged in two double layers of 16 fibers each, and read out at both sides by 64 MPPCs with an ad-hoc built readout electronics, was tested at the $\pi$M-1 line of the Paul Scherrer Institute (PSI) in Villigen, Switzerland. The detection efficiency and the trigger capability were measured on a beam containing protons, electrons, muons and pions with a momentum of 440 MeV/c. The measured average efficiency for protons for a double layer of scintillating fibers ($96.2 \pm 1.0 \% $) represents a guarantee of the good performance of this system as a trigger for the AMADEUS experiment.}


\begin{document}

\section{Introduction}

The AMADEUS experiment \cite{AMAD1,AMAD2} aims to perform low-energy charged kaons interactions in nuclear matter measurements, in particular to search for the so-called ``kaonic nuclear clusters''. The AMADEUS setup is going to be installed inside the KLOE detector \cite{KLOE} in the free space inside the drift chamber \cite{KLOEDC}. The experiment will then use the drift chamber and the calorimeter of the KLOE detector, together with a dedicated setup consisting of a target cell to be filled with deuterium, $^3He$ or $^4He$, and a dedicated trigger system, which will trigger on the back-to-back $K^+K^-$ pairs emitted from the decays of the $\Phi$ particles produced at the DA$\Phi$NE $e^+e^-$ collider of LNF-INFN \cite{DAF}.
For what concerns the trigger system, there are a series of constraints which have to be fulfilled:

\begin{itemize}

\item\emph{Reduced dimensions:} the system has to be small and compact, in order to fit inside the 50 cm diameter space inside the KLOE drift chamber.

\item\emph{Good performance in magnetic field:} the system has to work inside a 0.6 T magnetic field.

\item\emph{Work at room temperature:} the possibilities to install a cryogenic system inside the KLOE drift chamber are very limited, so the trigger detectors have to work at room temperature.

\item\emph{Detection efficiency:} a high detection efficiency of the system is fundamental in order not to lose good events.  

\item\emph{Good timing resolution:} in order to distinguish the slow kaons from background particles coming from beam losses by time of flight (TOF), a timing resolution of $\sim 300 \, ps \,(\sigma)$ is needed.

\item\emph{Trigger capability:} in addition to the timing information, the possibility to distiguish the kaons from charged MIPs (Minimum Ionizing Particles) by energy loss, enhances the performance of the system.
 
\end{itemize}

\noindent Small dimensions and the possibility to be operated in a strong magnetic fields are intrinsic properties of the Multixel Photon Counter detectors (MPPC) \cite{p1,p2,p3,p4}; these detectors can be easily coupled to plastic scintillating fibers, being used for the final readout of the photons created by the energy deposited by the kaons. Measurements of the time resolution and of the capacity to work at room temperature have been already performed and the results, fulfilling the above constraints, are published \cite{nim}. 
In this work, the trigger capability and the detection efficiency are investigated by using the $\pi$M-1 beam at the PSI.

\section{The trigger prototype setup}

\subsection{Detectors and fibers selection}

\begin{figure}[h]
\begin{center}
\mbox{\includegraphics[height=7.5cm]{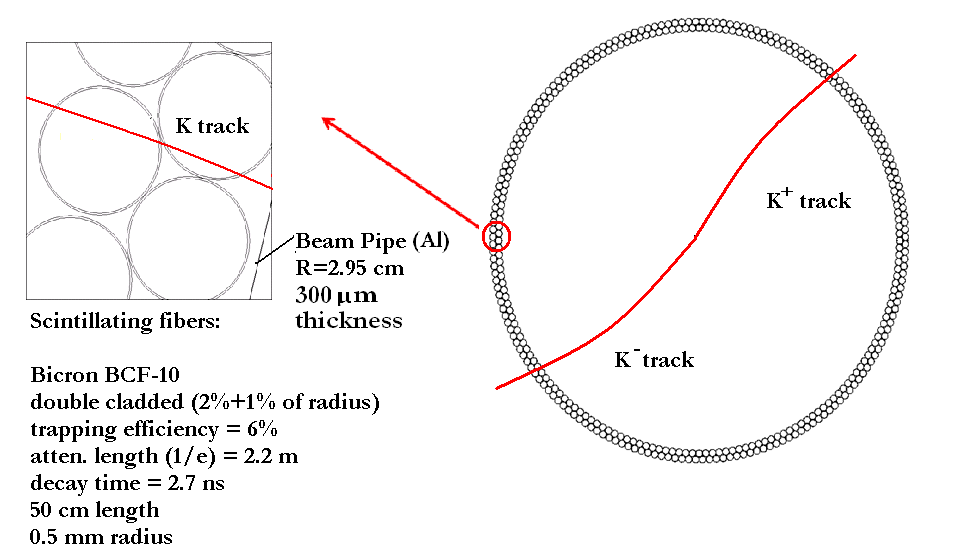}}
\caption{The double layer structure of the trigger system in the AMADEUS experiment surrounding the beam pipe, front view; a detail of the double layer geometry is shown in the left side. The $K^-$ and $K^+$ tracks are represented in red.}
\label{64chsetup}
\end{center}
\end{figure}

In an experiment like AMADEUS, the availability of a performant trigger system able to disentangle the signal, produced by kaons interacting in various targets, from background events, is extremely important for the success of the experiment.  The background is mostly generated by electrons and positrons (MIPs) lost from the circulating beams, either due to beam-gas interaction, or to interactions of particles in the same bunch, the so-called Touschek effect. The AMADEUS trigger system will consist in scintillating fibers which will be arranged in a cylindrical double layer structure surrounding the DA$\Phi$NE beam pipe (see fig. \ref{64chsetup}). Each fiber is going to be read at both ends by MPPC detectors. The two layers are positioned such as to increase the detection efficiency. Indeed, as confirmed by MC simulations, such a double layer geometry maximizes the probability that at least in one of the two layers the path of the kaon is long enough to generate a signal detectable by the MPPCs.
\newline In order to characterize the performances of such a trigger system, a prototype was built and tested. 
The prototype setup consists of two separable metallic rings hosting 32 scintillating fibers read at both ends by MPPCs. Each ring contains two double layers of 16 fibers each (8+8), reproducing the AMADEUS geometry. Each double layer is wrapped in black tape, in order to be light tight.
 
\noindent The choice of the MPPC device type was done taking into account a series of parameters, as the number of cells, the fill factor, the peak sensitivity wavelength (400 nm in our case), the photon detection efficiency, the dark current rate and the gain. After having performed tests on MPPC devices from different producers, we checked that those better matching our requirements are the HAMAMATSU S10362-11-050U ones, equipped with metallic package. These devices have 1 $mm^2$ effective active area, $1.25\times1.25\,mm^2$ chip size, 400 cells (50x50 $\mu m^2$ each), a fill factor of 61.5\%, 50\% PDE, 70\% quantum efficiency, a recommended operating voltage of $70\pm10\,V$, a maximum dark count rate of 800 KHz and a gain factor (at room temperature) of $7.5 \times 10^{5}$. 
For the scintillating fibers, the round shaped (1mm diameter) BICRON BCF-10 ones, 50 cm long were used.

\subsection{Readout electronics}

\begin{figure}[htbp]
\centering
\mbox{\includegraphics[width=12.cm]{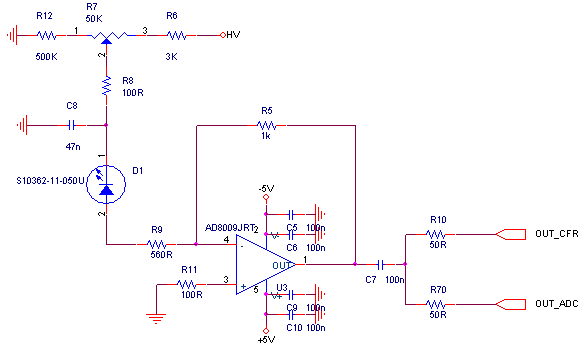}}
\caption{The electronic circuit of the preamplifier unit, consisting in an adjustable power supply and a 1 $K\Omega$ gain transimpedance amplifier.}
\label{ELEC1}
\end{figure}

Dedicated electronics modules were designed and built at LNF-INFN for the MPPC readout. 
MPPC signals are pre-amplified in a 8-channel board, providing a transimpedance of $1\,K\Omega$ and a $\sim \,10$ amplification factor. 
The analog output signals of these boards are individually processed by a constant fraction
discriminator module (also developed at LNF-INFN) which provides 64 ECL outputs and 5 NIM signals corresponding to the logic OR of the 64 channels. 
The electronic circuit of a single preamplifier unit is shown in fig. \ref{ELEC1}. 

\noindent Performances of the preamplifier and the time resolution obtained with this electronics have been checked in laboratory tests, using the a blue laser \cite{nim}.

\subsection{$\pi$M-1 beam settings}

The efficiency measurements were performed on the $\pi$M-1 line at the Paul Scherrer Institute (PSI) in Villigen \cite{PSI}; the beam can be tuned in order to have protons, with momenta in the range from 350 to 500 MeV/c, with a small contamination of MIPs ($\pi,e^-,\mu$). This configuration allows to test our setup in AMADEUS-like conditions. According to MC simulations, the energy loss in the 1mm thick scintillating fibers by the kaons coming from DA$\Phi$NE is comparable to the energy loss by 240 MeV/c momentum protons.
However, to reduce the MIPs contamination, the $\pi$M-1 proton beam was set to a momentum of 440 MeV/c; in this configuration, a residual MIPs contamination $\sim 24 \%$ was present. In tab. \ref{eloss} results of the simulations for energy loss in 1mm scintillating fibers are shown.

\begin{table}[htbp]
\begin{center}
\resizebox{0.6\textwidth}{!}{%
\begin{tabular}{|c|c|c|}
\hline
Particles	& Momentum ($MeV/c$)	&  Energy loss in 1 mm ($MeV$) \\
\hline
$K^-$  &  127   &  1.94 \\
\hline
$p$ &  240   &  1.92 \\
\hline
$p$ &  440   &  0.76 \\
\hline
\end{tabular}}
\end{center}
\caption{Energy loss in 1 mm thick scintillating fibers by DA$\Phi$NE kaons, 240 MeV/c and 440 MeV/c $\pi$M-1 protons calculated with a MC simulations (GEANT3).}
\label{eloss}
\end{table}

\begin{figure}[htbp]
\centering
\includegraphics[width=9.cm]{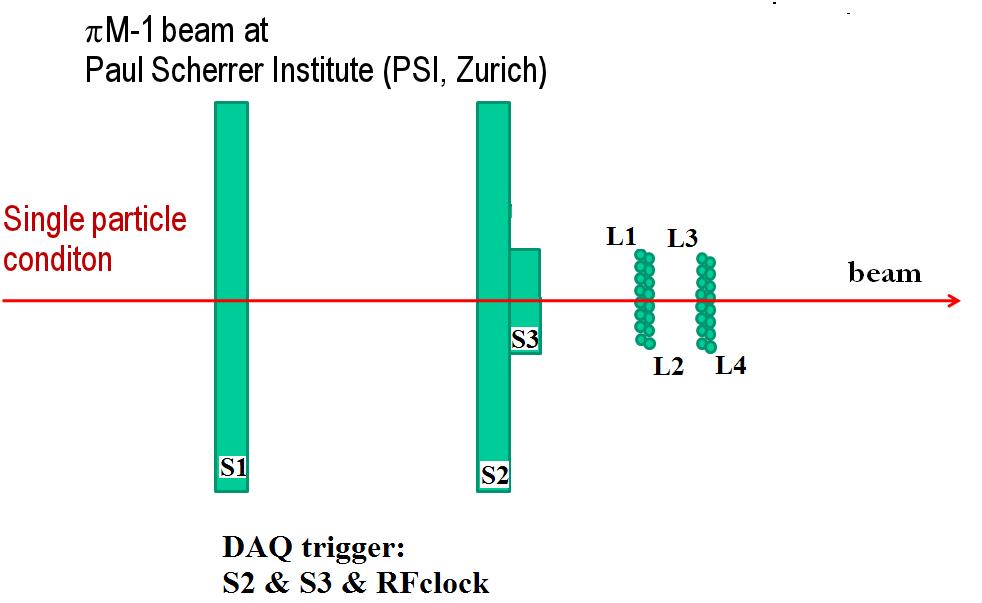}
\caption{Overall scheme of the test setup with the 3 scintillators and the two double layers scintillating fibers AMADEUS trigger prototype (not in scale).}
\label{whole}
\end{figure}

\begin{figure}[htbp]
\centering
\includegraphics[width=7.cm]{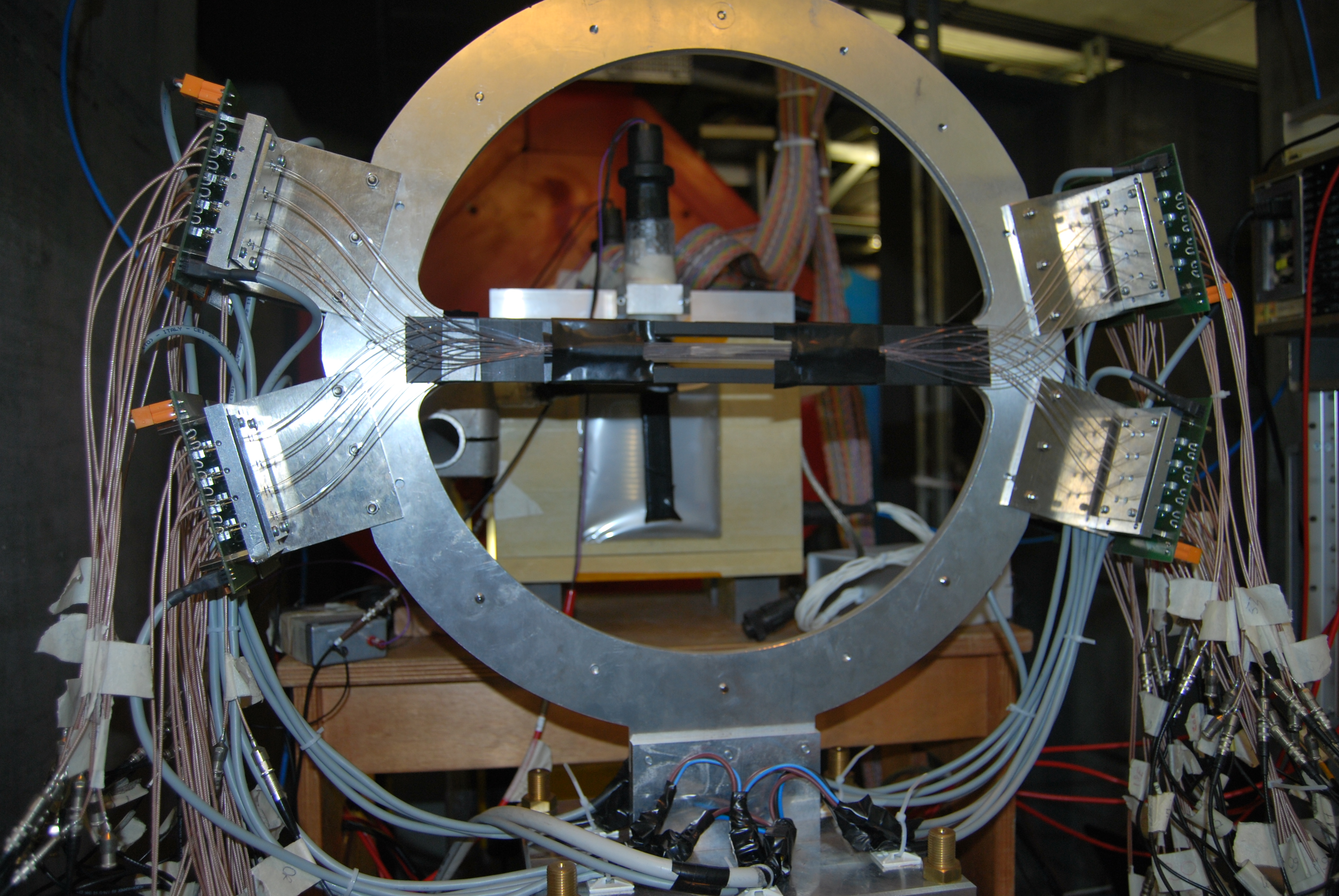}
\caption{Picture of the 64 channels prototype setup during the installation at PSI; the coupling of the fibers with the MPPCs is done through metallic holders where the fibers are inserted from one side untill touching the surface of the MPPC detector attached on the other side.}
\label{setup}
\end{figure}

\noindent The energy loss signal from 440 MeV/c momentum protons is a factor $\sim 2.5$ lower than the one expected in DA$\Phi$NE, which makes the conditions from PSI even more restrictive. \newline
In addition to this trigger prototype, 3 small scintillators were used to trigger the acquisition and for particle identification.
The scheme of the overall test setup (not in scale) as installed on the $\pi M-1$ beam line is shown in fig. \ref{whole}. Scintillators S1 and S2 have dimensions of $15\,cm \times 1.2\,cm, 1\,cm$ thickness, while S3 $15\,cm \times 2\,cm, 1\,cm$ thickness; the crossing area of S2 and S3 is then $2 \times 1.2 \,cm^2$.
In fig. \ref{whole} the arrow represents the proton beam; the trigger signal is given by S2 and S3, crossing at 90 degrees, in coincidence with the $\pi M-1$ RadioFrequency signal. Fiber layers are numbered from L1 to L4, while the fibers in each layer from 1 to 8. Using remote controlled adjustable slits, the beam was operated in single particle mode with a frequency of 50 MHz and a spot size on L4 of few centimeters \cite{PSI}. In fig. \ref{setup}, a picture of the prototype setup is shown.

\section{Experimental results}

The data from the 32 fibers and from the 3 scintillators were acquired with a VME DAQ chain, using two 32 channels QDCs (CAEN V792) and one 16 channel QDC (CAEN V792n) for the charge information, two 32 channels TDCs (CAEN V1190B) and one 16 channel TDC (CAEN V1190A) for the timing information, and a 32 channels ADC (CAEN V785) storing the temperature of the setup via a PT100 sensor.
The thresholds on the costant fractions were set to give an OR dark count rate of few Hz. The analyzed data correspond to a 690 minutes continuous run, performed in June 2012. 

\subsection{Events selection}

\begin{figure}[htbp]
\centering
\mbox{\includegraphics[width=7.5cm]{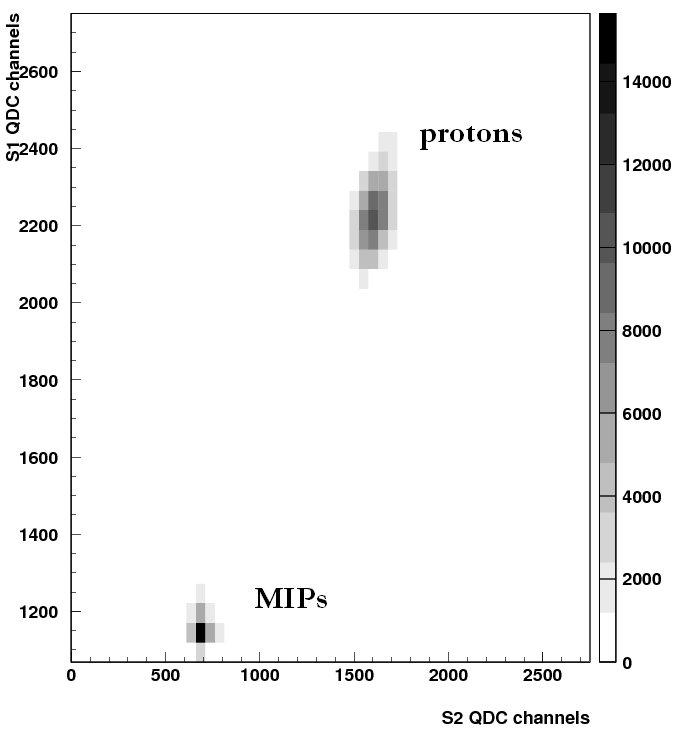}}
\caption{Correlation plot of S1 and S2 QDCs; protons and MIPs peaks are identified and labeled.}
\label{SELECTION}
\end{figure}

In the analysis procedure, the temperature stability ($\simeq $ 31 Celsius) was checked using a PT-100 sensor directly attached to the metallic holders of the MPPCs. Protons and MIPs are identified using the energy deposition correlation plot between S1 and S2, as shown in fig. \ref{SELECTION}. In order to have a clean subset of data, only events within one period of the RadioFrequency clock are considered.

\begin{figure}[htbp]
\centering
\mbox{\includegraphics[width=7.5cm]{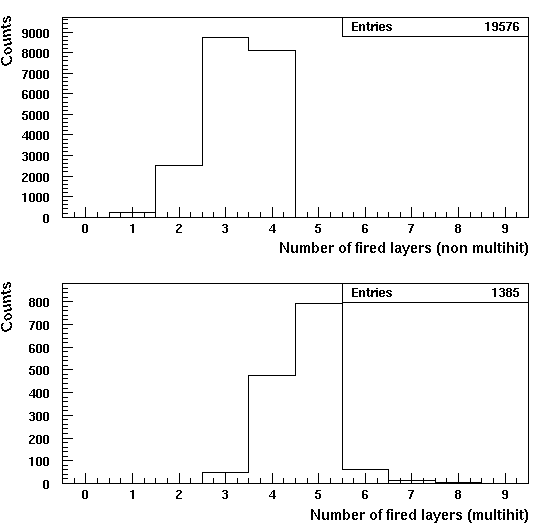}}
\caption{Single particle condition check; number of events in which 1, 2, 3, or 4 layers are fired when single hit per layer is request (up), and when at least one layer has a multihit (down).}
\label{SINGLE}
\end{figure}

\noindent A very important check relates to the protons multiplicity: working in single particle configuration is fundamental for a correct efficiency measurement, thus only events in which, for each layer, only one fiber is fired are kept for the efficiency determination.
In fig. \ref{SINGLE}, the total number of events in which 1, 2, 3 or 4 layers are fired when single hit per layer is requested (up) and when at least one layer has a double hit (down). The measurement shows that $\sim 93 \%$ of the events are single particle events; the rejected events, corresponding to the lower plot, contain also the cross talk between fibers of the same layer.

\subsection{Detection efficiency}

The detection efficiency of the double layer system measurement was performed using S1, S2 and L4 as reference. A "good event" is defined as an event in which a proton or a MIP, tagged by the two scintillators, is detected by one fiber of the L4.
The detection efficiency of the double layer (L1+L2) is then defined as the ratio of the number of events seen by a fiber in L1 or/and L2 Nr.Events(L1 AND/OR L2), with respect to the number of "good events" Nr.Events(L4):

\begin{equation}
Eff_{1+2} = \frac{Nr.Events(L1 \ AND/OR \ L2)}{Nr.Events(L4)}
\end{equation}

\begin{figure}[htbp]
\begin{center}
\mbox{\includegraphics[height=9.cm]{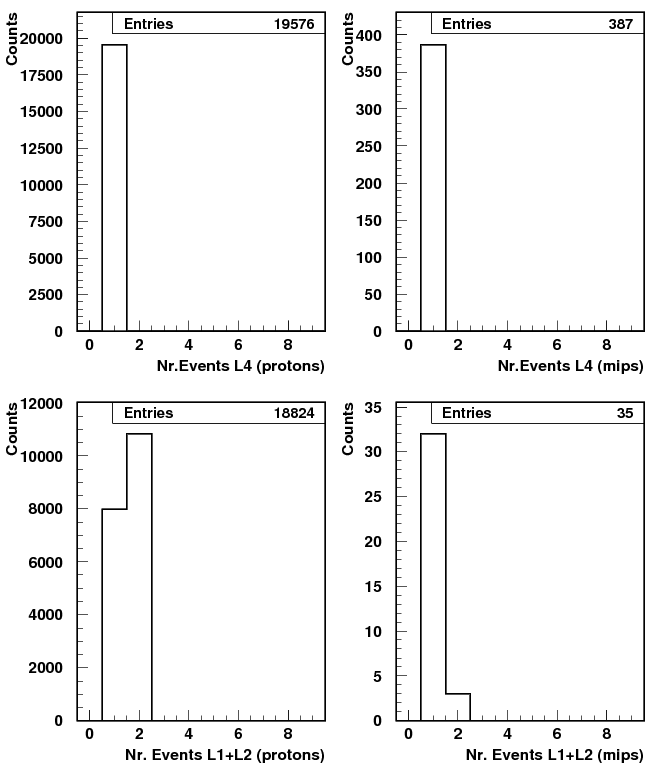}}
\caption{Total number of events with NHITS4=1 (up) and the ones detected by the double layer L1+L2 (down), for protons (left) and MIPs (right).}
\label{EFFTOT}
\end{center}
\end{figure}

\noindent Histograms of this measurement are shown in fig. \ref{EFFTOT}.

The measured efficiency for protons is:

\begin{equation} \label{wholeff}
Eff_{1+2}^{prot}=(96.2 \pm 1.0 (stat)) \% 
\end{equation}

\begin{figure}[htbp]
\centering
\mbox{\includegraphics[width=8.cm]{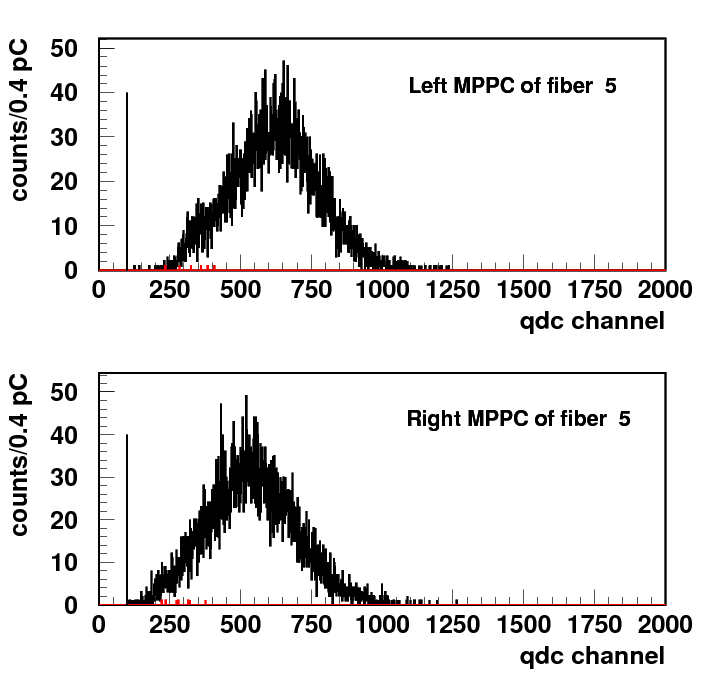}}
\caption{QDC plots of the two MPPCs reading the fiber number 5 of the L1; protons (black) and MIPs (red) events are shown. The vertical line represents the hardware threshold on constant fraction discriminators ($\simeq$ 10 photons).}
\label{QDC}
\end{figure}

For what concerns the MIPs, the measured value of $Eff_{1+2}^{MIPs}=(9.0 \pm 1.6(stat)) \% $ implies a rejection factor of more than $90 \%$. 
In addition to the double layer, the average efficiency for a single layer has been measured to be $Eff_{1lay}^{prot}=(75.5 \pm 0.8(stat)) \%$ and $Eff_{1lay}^{MIPs}=(5.9 \pm 1.2(stat)) \%$.
For a complete characterization of the prototype, very important information can be obtained from the charge deposition; as an example, the QDC plots of the two MPPCs reading the fiber 5 of L1 are shown in fig. \ref{QDC}, for protons (black) and MIPs (red). 

\noindent These plots show how, setting a hardware threshold on the discriminators corresponding to $\simeq\,10$ photons signal, more than $90\%$ of the MIPs can be rejected without losing "good events". It is important to observe that, as commented in section 2.3, the signal coming from the kaons at DA$\Phi$NE is a factor $\sim2.5$ bigger than the one corresponding to the PSI protons; this implies that the hardware threshold for MIPs rejection could be set to a higher value obtaining a bigger rejection factor. 
The same analysis has been done selecting each fiber, instead of the whole L4, as trigger for good events; results of this position scan of the detection efficiencies of (L1+L2) are shown in tab. \ref{effiber}. Due to the poor statistics of the MIPs events, this scan could only be done for the protons.

\begin{table}[htbp]
\begin{center}
\resizebox{0.45\textwidth}{!}{%
\begin{tabular}{|c|c|}
\hline
Triggering fiber in L4 & Efficiency of (L1+L2) (\%) \\
\hline
1  &  $91.1 \pm 1.3$   \\
\hline
2  &  $98.3 \pm 1.3$   \\
\hline
3  &  $97.2 \pm 1.3$   \\
\hline
4  &  $97.5 \pm 1.3$   \\
\hline
5  &  $99.1 \pm 1.1$   \\
\hline
6  &  $99.6 \pm 1.6$   \\
\hline
7  &  $98.8 \pm 1.0$   \\
\hline
8  &  $93.2 \pm 1.0$   \\
\hline
\end{tabular}}
\end{center}
\caption{Position scan of the detection efficieny. Each fiber is individually used as trigger for good events and the corresponding events on the double layer are counted.}
\label{effiber}
\end{table}

The detection efficiencies for all fibers are compatible whithin each other, except for the two external ones (1 and 8); this small difference is due to edge effects of the fibers layer. 

\section{Conclusions}

The detection efficiency and the trigger capability of the trigger prototype for the AMADEUS experiment, consisting of a double layer of scintillating fibers read by MPPCs, have been investigated at the $\pi M$-1 beam at PSI. 
The detection efficiency for protons was measured to be $Eff_{1+2}^{prot} = (96.2 \pm 1.0) \%$, with a capacity to reject the MIPs of ($91.0 \pm 1.6) \%$. This latter result can be further improved, considering that in the AMADEUS experiment the coincidence of two signals coming from a $K^+K^-$ pair is requested and that the expected energy loss by the kaons in DA$\Phi$NE in the fibers is a factor $\sim2.5$ bigger than the one of the protons of the PSI $\pi M-1$ beam. 
The results reported in this paper, taken together with the very good timing resolution of the system, show that the technical solution considered for the AMADEUS trigger system fulfills the requirements of the experiment and will allow to perform unprecedented measurements relevant to the low-energy QDC in the strangeness sector.

\acknowledgments

Part of this work was supported by the European Community-Research
Infrastructure Integrating Activity ``Study of Strongly Interacting Matter''
(HadronPhysics2(HP2), Grant Agreement No. 227431) and HadronPhysics3 (HP3), Contract No. 283286 under the Seventh Framework Programme of EU. 
We thank to the PSI staff and in particular to Dr. Konrad Deiters and the whole
$\pi$M-1 line team  for the excellent working conditions and for their support.


\begin{thebibliography}{9}


\bibitem{AMAD1} AMADEUS collaboration, \emph{AMADEUS phase-1: physics, setup and roll-in proposal, LNF internal release} {\bfseries LNF--07/24(IR)} (2007).
\bibitem{AMAD2} AMADEUS collaboration, \emph{Towards exclusive antikaonic nuclear cluster search with AMADEUS, Nucl. Phys. A} {\bfseries 804} (2008) 286.
\bibitem{KLOE} KLOE and KLOE-2 Collaborations, \emph{KLOE results in kaon physics and prospects for KLOE-2, Nucl. Phys. Proc. Suppl. B} {\bfseries 249} (2012) 225.
\bibitem{KLOEDC} M. Adinolfi et al., \emph{The Kloe Drift Chamber, Nucl. Inst. and Meth. in Phys. Res. A} {\bfseries 461} (2001) 25. 
\bibitem{DAF}  M. Boscolo et al., \emph{Luminosity and background measurements at the $e^+e^−$ collider upgraded with the crab waist scheme, Nucl. Inst. Meth. A} {\bfseries 621} (2010) 121.
\bibitem{p1} Z. Sadygov et al., \emph{Three advanced designs of micro-pixel avalanche photodiodes: Their present status, maximum possibilities and limitations, Nuclear Instruments and Methods in Physics Research A} {\bfseries 567} (2006) 70.
\bibitem{p2} N. Otte, \emph{The Silicon Photomultiplier - A new device for High Energy Physics, Astroparticle Physics, Industrial and Medical Applications} SNIC Symposium, Stanford, California, 3-6 April 2006.
\bibitem{p3} M. Dinu et al., \emph{Development of the first prototypes of Silicon PhotoMultiplier (SiPM) at ITC-irst, Nucl. Inst. and Meth. in Phys. Res. A} {\bfseries 572} (2007) 422.
\bibitem{p4} M. McClish et al., \emph{Characterization and scintillation studies of a solid-state photomultiplier, Nucl. Inst. and Meth. in Phys. Res. A} {\bfseries 572} (2007) 1065.
\bibitem{nim} M. Bazzi et al., \emph{Experimental tests of the trigger prototype for the AMADEUS experiment based on Sci-Fi read by MPPC, Nucl. Instr. and Meth. in Pyhis. Res. A} {\bfseries 671} (2012) 125. 
\bibitem{PSI} $http://aea.web.psi.ch/beam2lines/beam\_pim1.html$.

\end{thebibliography}
\end{document}